\documentclass[a4paper]{jpconf}
\usepackage{graphicx}
\usepackage{amsmath}

\newcommand{\av}[1]{\langle #1 \rangle}\begin{document}
\title{\centerline{Theoretical status of $B \to K^* \mu^+\mu^-$:}\\  \centerline{The path towards New Physics} \medskip}
\begin{center}
{\sc S\'ebastien Descotes-Genon}\\[2mm]
{\em\small
Laboratoire de Physique Th\'eorique, CNRS/Univ. Paris-Sud 11 (UMR 8627)\\ 91405 Orsay Cedex, France
}
\end{center}

\begin{center}
{\sc  Lars Hofer, 
Joaquim Matias\footnote{Speaker}}\\[2mm]
{\em \small
Universitat Aut\`onoma de Barcelona, 08193 Bellaterra, Barcelona
}
\end{center}

\begin{center}
{\sc  Javier Virto}\\[2mm]
{\em \small Theoretische Physik 1, Naturwissenschaftlich-Technische Fakult\"at,\\
Universit\"at Siegen, 57068 Siegen, Germany
}
\end{center}




\ead{matias@ifae.es}

\begin{abstract}
We summarize  the status of the analysis of the 4-body angular distribution $B \to K^*(\to K \pi) \mu^+\mu^-$ using a basis of optimized observables. We provide a New Physics interpretation 
 of the pattern of deviations observed in the 1fb$^{-1}$ dataset  with the coefficient of the semileptonic operator $O_9$ being the main responsible for this pattern. Also  other  scenarios  involving more Wilson coefficients are briefly discussed. Further, we present
a detailed description of each of the hadronic uncertainties entering our predictions and suggest possible tests for alternative hadronic explanations. 
Our most accurate SM predictions for the optimized observables in various bins including all  uncertainties are also provided.

\end{abstract}

\section{Introduction}
\medskip

After the long awaited discovery of the Higgs particle, it was assumed that this new scalar particle  would be instrumental as a handle that could open a portal for New Physics (NP). Unfortunately, it was found that this scalar  resembles very much the SM Higgs particle with SM-like couplings up to the present precision. In this sense it will possibly be a long-term task to find some clear evidence of  NP in this direction. In the meanwhile, other portals for NP, like those  constructed upon rare B decays can open new directions in these searches. Recently, it was found that one of those rare B decays,  $B \to K^* (\to K \pi) \mu^+\mu^-$, seems particularly promising. Its full 4-body angular distribution  provides sensitivity to many different Wilson coefficients, of electromagnetic, semileptonic and scalar operators (including their chiral counterparts). It thus will help to draw a picture of the flavour sector of the fundamental theory that lies beyond the SM.  A large variety of angular observables \cite{9907386,0502060, 0805.2525, 0811.1214, 1006.5013,1105.0376,1106.3283, 1202.4266, 1207.2753, 1303.5794}
will  allow to test some of the above-mentioned Wilson coefficients with an unprecedented precision. 

The keypoint in this type of searches for NP is  to find the cleanest possible procedure to extract the interesting short distance electroweak information disentangling it from any possible QCD polluting source. 
At leading order (LO) one can classify observables in form factor independent (FFI)  and form factor dependent ones (FFD).
 Some time ago,  a procedure \cite{1202.4266} was proposed to factorize the information contained in the 4-body angular distribution  into products of a set of FFI observables  times FFD ones like the longitudinal polarization fraction ($F_{L}$).  The short distance information can then be extracted from the less QCD-polluted FFI observables, which were named clean or optimized observables $P_i^{(\prime)}$ \cite{1207.2753,1303.5794}. The first of them, named $P_1$ or $A_T^2$ \cite{0502060}, was designed to detect the presence of right-handed currents (RHC). Soon after, other observables followed, like $A_T^{re}$ \cite{1106.3283} (or $P_2$ \cite{1202.4266}) that correspond to the clean version of $A_{FB}$, and $P_3$  sensitive to the presence of NP weak phases. The set was completed by the $P_i^\prime$ observables with $i=4,5,6,8$. Particularly interesting are the observables $P_4^\prime$ and $P_5^\prime$ that represent the interference between the longitudinal amplitude of the $K^*$ with the parallel and the perpendicular one, respectively.
  In these last observables at low $q^2$ ($q^2$ is the square of the invariant mass of the dilepton pair) but in the region above 5-6 GeV$^2$, the semileptonic operators dominate opening an excellent window to test them. In parallel, the observable $P_2$ testing the interference between perpendicular and parallel amplitudes, exhibits a window of sensitivity to these semileptonic coefficients in a different region, namely at the position of its maximum around 2 GeV$^2$\cite{1502.00920} and at the position of its zero near 4 GeV$^2$. 

In 2013  in a ground-breaking  effort going beyond traditional  analyses (studying only $F_L$, $A_{FB}$ and the branching ratio) and  borrowing techniques from different fields, LHCb presented the first analysis of this set of optimized observables \cite{lhcb}. A first interpretation of this measurement presented in \cite{understanding}  pointed out a {\it  coherent pattern} of small and large tensions in different observables, mainly in $P_2$ (and $A_{FB}$) and $P_5^\prime$ that could be explained under the hypothesis that the coefficient $C_9$ of the semileptonic operator and possibly also the coefficient $C_7$ of the electromagnetic operator receive a NP contribution. 

The main outcome of the analysis in Ref.\cite{understanding} based on the data from the observables  
\begin{itemize} \item[i)] $P_{1,2}$, $P_{4,5,6,8}^\prime$, $A_{FB}$ of  $B \to K^* \mu^+\mu^-$ in the  $q^2$ bins [0.1,2], [2,4.3], [4.3,8.68], [14.18,16], [16,19], 
\item[ii)] radiative and dileptonic B decays:  ${\cal B}_{B \to X_s \gamma}$, ${\cal B}_{B \to X_s \mu^+\mu^-}$, ${\cal B}_{B_s \to \mu^+\mu^-}$, $A_I(B \to K^* \gamma)$, 
\end{itemize}
was to find  that the deviations from the SM values of the Wilson coefficients, $C_i^{\rm NP}=C_i-C_i^{\rm SM}$, at 68.3$\%$ C.L. lie within the following ranges:
\begin{eqnarray} \label{pat}
C_9^{\rm NP}&\in&[-1.6,-0.9], \quad\quad\quad\;\, C_7^{\rm NP}\in[-0.05,-0.01], \quad\quad\quad 
C_{10}^{\rm NP}\in[-0.4,1.0], \nonumber\\ 
C_9^{\prime }&\in&[-0.2,0.8], \quad\quad\quad\quad C_7^{\prime \rm NP}\in[-0.04,0.02],\quad\quad\quad\;\; C_{10}^{\prime }\in[-0.4,0.4]. 
\end{eqnarray}
Here, $C_9^{\rm NP}$ is consistent with zero (SM) only above 3$\sigma$, $C_7^{\rm NP}$ is consistent with the SM at 2$\sigma$, while the rest are consistent with the SM already at 1$\sigma$. In this solution one observes that $C_9^{\rm NP}+C_9^\prime<0$, $C_7^{\rm NP}<0$  and $C_{10}^{\rm NP}$ seems to show a mild preference for positive values due to ${\cal B}_{Bs \to \mu^+\mu^-}$. Remarkably, different groups \cite{1308.1501,1310.2478,1310.3887} using different observables, techniques and statistical approaches also confirmed the large contribution to $C_9^{\rm NP}$ of order $25\%$ with respect to the SM value. It was also claimed by some groups \cite{1308.1501,Hambrock:2013zya} using, for instance, the low-recoil 1fb$^{-1}$ data on $B^+ \to K^+\mu^+\mu^-$ \cite{Aaij:2012vr} that at the same time also the chirally flipped semileptonic coefficient was getting a large NP contribution such that $C_9^\prime + C_9^{\rm NP} \simeq 0$ in order to have a SM-like BR for the $B^+ \to K^+ \mu^+\mu^-$ mode. However, this scenario was in conflict  \cite{understanding,1311.3876}  with the anomaly in $P_5^\prime$ (4$\sigma$ deviation in the third bin of $P_5^\prime$) that required $C_9^\prime$ to be zero or  small such that $C_9^{\rm NP}+ C_9^\prime<0$ in order not to increase this anomaly. Fortunately an updated result from LHCb with 3fb$^{-1}$ data on $B^+ \to K^+ \mu^+\mu^-$ and $B^0 \to K^0 \mu^+\mu^-$ \cite{Aaij:2014pli} has solved this tension since both modes point  in the low-q$^2$ and large-q$^2$ region to the solution $C_9^{\rm NP}+C_9^{\prime}<0$  in  agreement with the anomaly in $P_5^\prime$.  

In summary, the present situation exhibits a coherent pattern of deviations in $B \to K^*\mu^+\mu^-$, $B^+ \to K^+ \mu^+\mu^-$, $B^0 \to K^0 \mu^+\mu^-$   but also $B_s \to \phi \mu^+\mu^-$ and the radiative and dileptonic observables mentioned above, pointing to the semileptonic coefficient $C_9$ as the main responsible, with all other coefficients not deviating significantly from zero and playing a less important role. It is important to emphasize already at this point that this conclusion obviously does not preclude the possibility that the new data can switch on with enough significance other Wilson coefficients.  

Our main goal here will be to discuss all new corrections that we have included in our predictions \cite{1407.8526} and in parallel analyse  
alternative explanations that has been raised to explain some of the deviations suggesting possible tests or cross checks to discriminate them from a NP solution. 

 The outline of the proceedings is the following: in Section 2 we discuss the theoretical framework at low $q^2$ where the strongest deviations from the SM have been found by LHCb. In Section 3 we detail  our method to include 
 factorizable and non-factorizable (including charm-loop) corrections. We also suggest  some tests to check the presence of huge charm-loop contributions and comment on some of the possible consistency tests that can be done on data. Section 4 is devoted to S-wave pollution. In Section 5 we comment on and explore other  scenarios. Finally we conclude in Section 6. We provide in an appendix our most accurate predictions  for the optimized basis of observables in various bins and in two different parametrizations including all corrections discussed before based on the method developed in \cite{1407.8526}.
 

\section{Theoretical Framework at low-$q^2$}
\medskip

In this section we will focus on the theoretical framework used to describe the large-recoil region  where the strongest deviations have been observed. Still for completeness we will also comment at the end  of this section on some aspects concerning the size of possible uncertainties that might affect either the large or the low recoil region.

A keypoint in the evaluation of the angular observables is to keep track of the correlations among form factors. In the large-recoil region there are basically two approaches:
\begin{itemize}
\item ``Improved QCDF approach": In this framework \cite{1303.5794}  the large-recoil symmetries between form factors are used to implement the dominant correlations among them. This is a transparent and general approach that is easy to cross-check and valid for any form factor parametrization (e.g. for the light-cone sum rules parametrizations \cite{KMPW} or \cite{bz}). The symmetries allow  the 7 form factors to be written in terms of only  two, so called  soft form factors $\xi_{\perp,\|}$ \cite{9812358}:
\begin{eqnarray} \label{correlations} 
\frac{m_B}{m_B+m_{K^*}}{V(q^2)} = \frac{m_B+m_{K^*}}{2E}{A_1(q^2)}
={T_1(q^2)} = \frac{m_B}{2 E} {T_2(q^2)} =  {\xi_\bot(E)}, \nonumber \\[2mm]
\frac{m_{K^*}}{E} {A_0(q^2)} =
\frac{m_B+m_{K^*}}{2E} {A_1(q^2)} - \frac{m_B-m_{K^*}}{m_B}{A_2(q^2)}
= \frac{m_B}{2E} {T_2(q^2)} - {T_3(q^2)} = {\xi_\|(E)}. 
\end{eqnarray}
Then the soft form factors can be computed in a specific parametrization. 
In \cite{1407.8526} we employed this approach considering all symmetry breaking corrections  to the relations Eq.(\ref{correlations}). Our predictions take into account factorizable and non-factorizable $\alpha_s$ corrections computed within QCDF \cite{0412400,0008255,0106067}, as well as  factorizable and non-factorizable power corrections, discussed in Sec.3,  including  charm-loop effects. The analytic structure of the large-recoil correlations led to the construction of the FFI observables  $P_i$ and $P_i^{CP}$ \cite{0502060,1202.4266,1207.2753,1303.5794} that exhibit a sensitivity to the soft form factor suppressed in $\alpha_s$ or $\Lambda/m_b$.
All bins  are traditionally considered in our analysis.

\item ``Ball-Zwicky Form Factor approach": Here  a specific set of full form factors (Ball-Zwicky \cite{bz}) is used. Factorizable $\alpha_s$ and factorizable power corrections are automatically included  with correlations associated to this particular parametrization.  All other corrections have to be included and/or estimated exactly as in the previous approach, non-factorizable $\alpha_s$ correction taken from QCDF (here also soft form factors are necessary), non-factorizable power corrections (see \cite{straubnew}) and charm-loop effects. This approach has been employed in \cite{1308.1501} where
the basis $S_i$ and $A_i$ \cite{buras} is  used and only bins below 6 GeV$^2$ are considered.

\end{itemize}

Both approaches are useful and complementary, should converge and give comparable results and error sizes. Of course, both methods can be used to compute both type of observables $P_i$ and $S_i$. For a comparison one can apply the general first approach (QCDF) to the particular form factor parametrization (BZ) used in the second one and the observables studied in the second approach \cite{straubnew}. We will show an example of this comparison in Sec. 3.1.


In the low-$q^2$ region, where we focused  the previous discussion, the large set of observables measured provide a nice cross-check of the observed deviations. 
For this reason we think it is important to consider all available bins. 
Sometimes it is stated that in bins beyond 6 GeV$^2$ soft-gluon corrections to the virtual charm-loop could have a large impact. 
However, there are two important remarks: first, the partial soft-gluon charm correction computed in Ref. \cite{KMPW}  is found to be positive implying that its effect is to enlarge and not reduce the anomaly, requiring an even larger NP contribution to compensate for it. Second, it is manifest in the plots  of  Fig. 5 and 7 of Ref. \cite{KMPW} that the impact is moderate up to 8 GeV$^2$.  For some amplitudes ($M_3$) it is even smaller than the impact at very low $q^2$. In our predictions  we take this calculation as an estimate of the effect and allow both signs for this correction as explained later on. In any case it is advisably to cut the experimental data at 8 GeV$^2$ where the impact of the charm-loop contribution calculated in \cite{KMPW} is still  moderate.




In the large-$q^2$  region, one of the main sources of uncertainty and model dependence  is related to the treatment of the  observed set of resonances  by LHCb in the data of the partner channel $B^+ \to K^+ \mu^+\mu^-$.  This observation prevents one from taking small bins afflicted by the resonance structures. In \cite{buchalla} a quantitative estimate of duality violation is given. Unavoidably, one needs to use a model for this estimate, still the result is that the integrated single low recoil bin gets a duality-violation impact of a few percent (5 $\%$ in \cite{pirjol} or 2$\%$ in \cite{buchalla}). On top of that, there is  certainly an impact from the choice of the starting point for the integrated bin that might coincidence with the position of a resonance. It is thus under debate how large this violation should be taken and where to start the long bin. In \cite{buchalla} the  estimate for $B \to K\mu^+\mu^-$  is given for  $q^2 \geq  15$ GeV$^2$ while first data on $B \to K^* \mu^+\mu^-$, where a similar problem will arise, starts at 14.18 GeV$^2$. An interesting exercise to show how sensitive and dependent on the starting point the result is,  was done in \cite{1310.3887}. There lattice form factors were used to  analyze  the low-recoil region for $B \to K^* \mu^+\mu^-$ and $B_s \to \phi \mu^+\mu^-$. It was found that if the two bins were taken the best fit value was given by a solution with $C_9^{NP}$ negative and $C_9^\prime$ positive (both with a significance slightly above 1 $\sigma$), while if only the last bin was used (starting at 16 GeV$^2$) the result on $C_9^{NP}$ was practically unaffected while $C_9^\prime$ became consistent with zero. In this sense again it would be interesting that the new data on $B \to K^* \mu^+\mu^-$ starts at 15 GeV$^2$ or above as it has been done for the 3 fb$^{-1}$ data of $B \to K \mu^+\mu^-$.

\section{Methodology beyond QCDF and alternative explanations}
\medskip

In this section we discuss in detail  the method to include factorizable and non-factorizable power corrections (including charm-loops)  that we used for our predictions\cite{1407.8526}. We also provide some tests to  show the robustness of the predictions and  point out possible ways to check, using the new data, some alternative explanations (huge charm loops) proposed to explain one of the deviations. We also discuss model independent relations between observables that allow to perform clear tests on the data, assuming the absence of new CP violating phases beyond the SM.

\subsection{\bf Factorizable power corrections}
\medskip

The decomposition of a full form factor contains in general three pieces,

\begin{equation}F(q^2)=  F^{\rm soft}(\xi_{\perp,\|}(q^2))  + \Delta F^{\alpha_s}(q^2) + \Delta F^{\Lambda}(q^2),\label{decomposition}\end{equation} where the first piece includes the soft form factors that implement automatically the dominant correlations among form factor given in Eq.(\ref{correlations}), $\Delta F^{\alpha_s}(q^2)$ represents the known QCDF correction induced by hard gluons, and $\Delta F^{\Lambda}(q^2)$ are the factorizable power corrections of  ${\cal O}(\Lambda/m_b)$. Following \cite{jaeger} we parametrize the latter correction, which is expected to be small compared to hadronic uncertainties, as an expansion in $q^2$:
\begin{equation}\Delta F^{\Lambda}(q^2)= a_F + b_F \frac{q^2}{m_B^2} + c_F \frac{q^4}{m_B^4}+...\label{param}\end{equation}

The decomposition (\ref{decomposition}) is not unique, one can always reshuffle the contributions among the different terms. It is necessary to fix a renormalization scheme, namely define $\xi_{\perp,\|}$ in terms of full form factors. Our implementation of this correction \cite{1407.8526} makes particularly emphasis on two aspects. The first aspect is the importance of respecting the correlations among {\it all} the parameters of Eq.(\ref{param}) of different form factors.  There are exact kinematical relations at $q^2=0$, like $T_1(0)=T_2(0)$ implying $a_{T1}=a_{T2}$, but also correlations among the $a_{F_i}$, $b_{F_i}$ and $c_{F_i}$ coefficients that originate from the definition  of the soft form factors in terms of the full form factors.  For this reason it is very important to always work consistently within a form factor parametrization at a time not mixing them. The second  aspect is the relevance of choosing the most appropriate scheme. The choice of scheme determines which part of the ${\cal O}(\alpha_s,\Lambda/m_b)$ correction
will remain in the function $\Delta F^{\alpha_s}$ or $\Delta F^{\Lambda}$ and which part will be absorbed in to the $F^{\rm soft}(\xi_{\perp,\|}(q^2))$. While the  perturbative $\Delta F^{\alpha_s}$ correction can be computed explicitly in each scheme,  the lack of a full control of the correlations among the errors of $\Delta F^{\Lambda}$ induces a scheme dependence at order ${\cal O}(\Lambda/m_b)$ on the otherwise scheme independent observables. In this situation a proper choice of scheme is mandatory  in such a way that the effects of unknown power corrections get absorbed as much as possible into the soft form factors in order not to artificially inflate uncertainties . A simple example  illustrates this point: assume an observable that only depends on a single form factor $F_1$. A good scheme would be  the one that takes this form factor directly as input, so that all power corrections are absorbed and would not appear from the beginning. A bad scheme would fix as input another form factor $F_2$, and  in the process of relating it to $F_1$ unknown power corrections would be introduced increasing the uncertainty of the result artificially. Our choice to determine the soft form factor  $\xi_{\perp}(q^2)$ is given by \begin{equation}
  \xi_{\perp}(q^2)\,\equiv\,\frac{m_B}{m_B+m_{K^*}}V(q^2).
  \label{eq:xiV}
\end{equation}
%
This definition eliminates all corrections to the form factor $V$ leading to $\Delta V^{\alpha_s}(q^2)=\Delta V^{\Lambda}(q^2)=0$. A different choice consists in defining $\xi_{\perp}(q^2)$ from $T_1$ like in \cite{jaeger} leads to larger uncertainties in $P_5^\prime$.

The soft form factor $\xi_{\parallel}$ can be defined as
\begin{equation}
  \xi_{\parallel}(q^2)\,\equiv\,\frac{m_B+m_{K^*}}{2E}A_1(q^2)\,-\,\frac{m_B-m_{K^*}}{m_B}A_2(q^2),
  \label{eq:xiA12}
\end{equation}
as done for example in Ref.~\cite{0412400}. This definition
minimizes power corrections in the form factors $A_{1,2}$ by correlating $\Delta A_1^{\alpha_s}(q^2),\Delta A_1^{\Lambda}(q^2)$
with $\Delta A_2^{\alpha_s}(q^2)$ and $\Delta A_2^{\Lambda}(q^2)$. Another possible determination of this form factor uses the form factor $A_0(q^2)$ \cite{jaeger}, albeit perfectly acceptable as a choice  this FF does not enter by construction any of the optimized observables. 

In our computation we first determine the  parameters $a_F,b_F$ and $c_F$ such that the central value of the full form factor is exactly reproduced. We perform a fit to these parameters and keep the  correlated (non-zero) central values obtained: ${\hat a_F}, {\hat b_F}, {\hat c_F}$. 
We have applied this procedure to two different form factor determinations KMPW \cite{KMPW} and BZ \cite{bz}, taking always one at each time to respect correlations. This induces a shift in the central values as compared to the previous predictions in \cite{understanding}.
For the errors associated to the factorizable power corrections we take $\Delta F^{\Lambda} \sim F \times {\cal O}(\Lambda/m_b) \sim 0.1 F$ which amounts to assign an error of order 100$\%$ with respect to  the fitted central values ${\hat a_F}, {\hat b_F}$ and ${\hat c_F}$. Finally, we vary the parameters within the range ${\hat a_F}-\Delta {\hat a_F} \leq a_F \leq {\hat a_F} +\Delta {\hat a_F}$ and similarly for $b_F$ and $c_F$. Note that our error estimate for the factorizable power corrections is solely based on dimensional arguments and exhibits a marginal dependence on the form factor parametrization used.

Finally to close this section on factorizable power corrections we compare  the results obtained using the two different approaches (QCDF versus Ball-Zwicky FF)\footnote{We will not include charm-loop effects in this comparison in our prediction because we only computed them for KMPW and not for BZ. In the same way we will not take it into account in \cite{jaeger}} . Let us take, for instance, the bin [2,4.3] for the observable $S_5$ using the improved QCDF approach for the particular case of the BZ parametrization. Our prediction  from \cite{1407.8526} for this observable is $S_{5 \, BZ}^{[2,4.3]}=-0.17 \pm 0.04$. Now we  compare this prediction  with the one in the recent paper \cite{straubnew} obtained using the different method based on the correlated  BZ form factors. The result there is $S_5^{[2,4.3]}=-0.17 \pm 0.04$ exactly the same.\footnote{Notice that if we use the KMPW parametrization our prediction becomes $S_{5 \, KMPW}^{[2,4.3]}=-0.17^{+0.08}_{-0.06}$ where the difference comes as it is well known  from the larger hadronic form factor errors in KMPW, while the error associated with factorizable power corrections is similar in size as with BZ form factors. This is the parametrization we take for our predictions.} This perfect agreement is remarkable given  the  different treatment of the  error  associated to factorizable power corrections. 
However, if one does the same exercise using the value of $P_5^\prime$ in the same bin given in \cite{jaeger} and transforming it into a prediction for $S_5$ {\it assuming zero error} from $F_L$, one gets $S_{5 \, JC}^{[2,4.3]}=-0.11^{+0.13}_{-0.11}$   with the error being a factor of 3 larger  as compared to the predictions from  \cite{straubnew} and \cite{1407.8526}.  We have also done the same exercise for the [1-6] bin of $S_5$ with identical outcome (see \cite{talk} for further details). 
  
  \subsection{\bf Non-Factorizable power corrections}
\medskip  

Concerning non-factorizable contributions, we will separate the non-perturbative contributions appearing at subleading order in the $1/m_b$ expansion into two types. First we will single out the hadronic contribution not related to Wilson coefficients steaming from the  piece of the three hadronic  ${\cal T}_i(q^2)$ that parametrize the matrix element  $\langle K^* \gamma^*|H_{eff}|B\rangle$. We obtain it taking the limit ${\cal T}_i^{\rm had} = {\cal T}_i|_{C_7^{(\prime)}\to0}$.
Finally, we multiply each of these amplitudes, that will serve as a normalization, with a complex $q^2$-dependent factor
\begin{equation}
{\cal T}_i^{\rm had}\to \big(1+r_i(q^2)\big) {\cal T}_i^{\rm had},
\end{equation}
where
\begin{equation} r_i(s) = r_i^a e^{i\phi_i^a} + r_i^b e^{i\phi_i^b} (s/m_B^2) + r_i^c e^{i\phi_i^c} (s/m_B^2)^2. \end{equation}
We define our central values as the ones with $r_i(q^2)\equiv0$, and estimate the uncertainties from non-factorizable power corrections
by varying $r_i^{a,b,c}\in [0,0.1]$ and $\phi_i^{a,b,c} \in [-\pi,\pi]$ independently, corresponding to a
$\sim 10\%$ correction with an arbitrary phase. The uncertainties for each observable are then obtained by performing a random
scan and taking the maximum deviation from the central values to each side, to obtain  upward and
downward error bars. A separated discussion of the treatment of non-factorizable contributions coming from charm loops is given in next section.




\subsection{\bf Charm-loop pollution}
\medskip

Part of the $c{\bar c}$-loop contributions have been already included in  the non-factorizable contributions (hard-gluon exchange). The remaining long-distance contributions from  $c{\bar c}$ loops are still under debate. We will rely for these corrections on the  partial computation  \cite{KMPW}. It is important to remark that the soft-gluon contribution of \cite{KMPW} coming from 4-quark and penguin operators induces a {\it positive} contribution to $C_9^{\rm eff}$ whose effect is to enhance the anomaly. We are  interested only in the long-distance contribution $\delta C_9^{\rm LD}(q^2)$, so we subtract the  perturbative LO part  and include the shift due to a different reference value for $m_c$ (see \cite{1407.8526} for more details on the procedure used). In order to be  conservative, and in particular  given the discussion on the sign of this contribution, we use the result of \cite{KMPW}
as an order of magnitude estimate, including it as  
\begin{equation}
{\cal C}_{9} \to {\cal C}_{9} + s_i \delta C_9^{\rm LD}(q^2)\ ,
\end{equation}
with $s_i$ scanned in the range $[-1,1]$ individually for each amplitude $M_i$, with $i=1,2,3$ 
 (remember that the result of the calculation is $s_i=1$). 
 
In an attempt of explaining mainly $P_5^\prime$, the possibility of a huge charm contribution has been proposed \cite{zwicky}. As will be explained below the fundamental problem of this explanation is to focus on a single observable, 
 while it is important to consider  the global picture and the combined input from all observables that can put in serious difficulty this explanation. At a practical level this explanation transfers the responsibility of the deviation from $C_9^{\rm NP}$ to a modified  $Y(q^2)$ (with a prefactor) and adds a similar function (and prefactor) also to $C_9^\prime$. Here 
$Y(q^2)$ is the 4-quark contribution to the effective Wilson coefficient $C_9^{\rm eff}$. 
While $C_9^{\rm NP}$ is a global and constant effect entering all bins, the modified $Y(q^2)$  is a $q^2$  dependent  function that is expected to give a  small contribution around 2 GeV$^2$.
Several examples of the impact of an hypothetical parametrization for this  function were presented in \cite{zwicky}. All the illustrative examples shown there are disfavoured for some observable, either $P_2$, $P_5^\prime$ or $P_1$, this shows the importance of performing a global analysis.
 
It is precisely the question of a global versus a local $q^2$ dependent effect 
which discriminates between a NP scenario and a charm-loop effect.
 For instance, as explained in \cite{1502.00920} the position of the maximum of the observable $P_2$ is practically unaffected by a hypothetical large charm contribution if one takes into account the experimental constraint on the zero of $P_2$ (the same as $A_{FB}$). This implies that while the shift in the position of the zero of $P_2$ could be explained with this hypothetical huge contribution, this is not the case for the maximum.  On the contrary, a $C_9^{NP}$ contribution would shift in a precise way {\it both} the position of the maximum and the position of the zero of $P_2$. Present data  seems to prefer a shift of both maximum and zero in contrast to the hypothetical large charm contribution. If new data confirms the double shift, the claim about a huge charm-loop contribution will be challenged.


A second fundamental problem of  a large charm explanation is the recently measured
$R_K$ \cite{1406.6482}. It is clear that, given its universal character that does not distinguish muons from electrons,
the charm loop cannot explain $R_K$, while the same NP scenario that is favoured by the anomaly in $B \to K^*\mu^+\mu^-$ can explain it if NP couples only to muons \cite{ghosh}.  Finally,  the strong $q^2$ dependence of the charm loop proposed in \cite{zwicky} tends to generate more than one zero in the region of low $q^2$ for $P_2$ and $P_5^\prime$ (see plots of Fig.12 in \cite{zwicky}). If no second zero is found,  many of those solutions will be severely constrained. If  ${P_5^{\prime}}_{[6,8]}$ is  equal or above   ${P_5^{\prime}}_{[4,6]}$,  it would provide a positive test in favour of this alternative explanation, and an important negative test otherwise.

\subsection{\bf Large statistical fluctuation and  consistency tests on data}
\medskip

Another possible explanation of the strong deviation  in the third bin of $P_5^\prime$ is the possibility of a purely  statistical fluctuation or an isolated experimental problem in that bin. This is  possible and  very plausible if one decouples the third bin of $P_5^\prime$  from tensions in the rest of observables.
 The aim behind the paper \cite{nicola} was to explore if it is possible to establish a connection between different observables under reasonable and well defined  assumptions, but independently of any effective Hamiltonian computation, just relying on symmetry arguments of the distribution. If such a relation exists  and is confirmed by experimental data, the anomaly in $P_5^\prime$ cannot be considered as an isolated problem.

A simple but powerful relation obtained in \cite{nicola} under the assumption of real Wilson coefficients shows that

\begin{equation} [P_4^{\prime 2} + \beta^2 P_5^{\prime 2}]_{q^2=q_0^2}=1 + \eta(q_0^2), \end{equation}
where $q_0^2$ is the position of the zero of $P_2$ and $\eta(q_0^2) \sim 10^{-3}$ if no RHC are present. This  equation establishes a  non-trivial link between the zero of $P_2$ and the anomaly in $P_5^\prime$. If the position of the zero of $P_2$ is at a higher $q^2$ value  than predicted by the SM  (${q_0^{2}}_{SM}=4$ GeV$^2$), then 
taking into account that the measured $P_4^\prime$ has a strong positive  slope towards 1, the value of $P_5^\prime(q_0^2)$ should be close to 0.
This implies that  
the small value of $P_5^\prime$ in that region and the shift of the zero of $A_{FB}$ are two sides of the same coin. This establishes a strong link between two observables based on symmetries \cite{1202.4266,1005.0571}. 

Also the more general relation \cite{nicola} \begin{equation}\label{relation} P_2 = \frac{1}{2} \left[ P_4^\prime P_5^\prime + \frac{1}{\beta} \sqrt{(-1 + P_1 + P_4^{\prime 2}) (-1 -P_1 + \beta^2 P_5^{\prime 2})}\right], \end{equation}
together with exact bounds like \cite{nicola} \begin{equation} P_5^{\prime 2}-1 \leq P_1 \leq 1-P_4^{\prime 2},\label{bound}\end{equation} allows to perform several tests on the data. Let us briefly mention some of them:

\begin{itemize} 
\item[i)] A comparison between the measured value of $P_2$ and  the value obtained from Eq.(\ref{relation}) using the measured values of $P_1$ and $P_{4,5}^\prime$ show an excellent agreement for the second bin (see Fig.1). The third bin  shows a tension of 2.4$\sigma$ between the measured value of $P_2$ and the one obtained from Eq.(\ref{relation}). 
This implies that data on the third bin should change either in $P_2$ or in $P_1$, $P_{4,5}^\prime$ to get a better consistency between the measured and the obtained value from the other measurements.
\item[ii)]  The exact bound Eq.(\ref{bound}) implies that, if one writes $P_4^\prime =1 + \delta$ in the third large recoil  bin, then $P_1 \leq -2 \delta + {\cal O}(\delta^2)$. If $P_1$ is positive (as data suggest)  in that third bin, $\delta$ should be negative and $P_4^\prime$ be slightly below 1. This is consistent with data given the large error bars, although the higher positive values of $P_1$ would be in tension with that bound.
\item[iii)] Finally, the first low-recoil bin [14.18-16.00] GeV$^2$ shows a large discrepancy of 3.7$\sigma$ between the measured value for $P_2$ and the
value obtained from the data on $P_1$, $P_4^\prime$ and $P_5^\prime$. In this case, it is very plausible that the responsible of this disagreement can be traced to the value of $P_4^\prime$ in that bin.
\end{itemize}

A simple consequence for data in the third bin, pointed out in \cite{1407.8526} (see Appendix A3),   can be easily found by combining i) and ii). 
$P_2$ has a zero at $q_0^2$,implying that $P_2 = - \epsilon$ (with $\epsilon>0$) for $q^2 > q_0^2$, and from Eq.(\ref{relation}) it follows that 
\begin{equation} P_5^\prime < -2 \frac{\epsilon}{1+\delta} \label{order}. \end{equation}
This implies that one should naturally expects  $P_5^\prime$ to be below $P_2$ in that region, and this is the reason for the tension of 2.4$\sigma$ found in Fig.1. Consequently we expect that new data should rearrange the third bins of $P_2$ and $P_5^\prime$ making more manifest the ordering of these two observables. 




\begin{figure}
\begin{center}
\includegraphics[height=6cm,width=8cm]{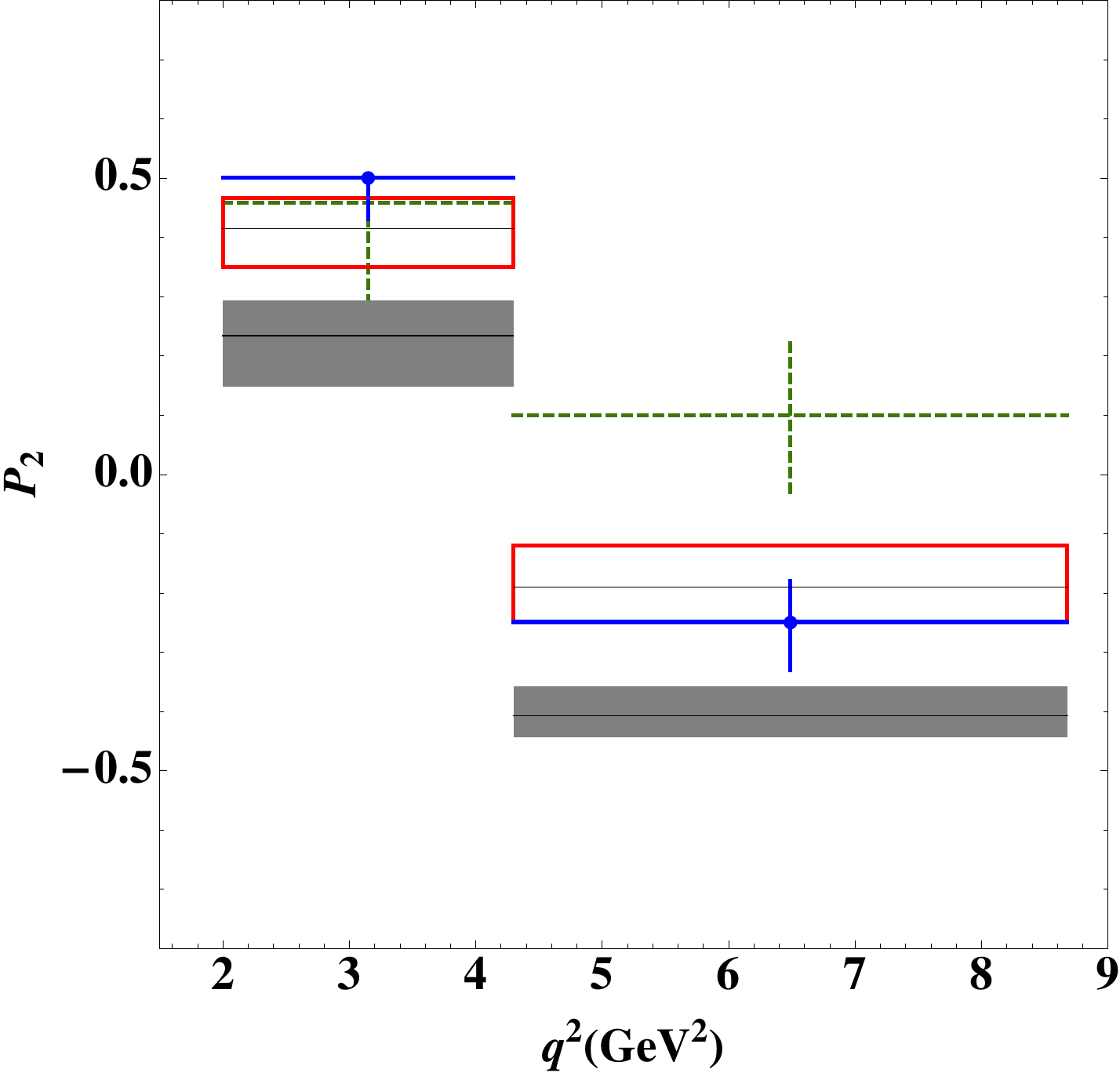}
\end{center}
\caption{\label{label} Comparison between the measured value of $P_2$ (blue cross) and the value obtained from the measurements of $P_1$, $P_{4,5}^\prime$ using Eq.(\ref{relation}) (green cross). Gray band is SM.}
\end{figure}




\section{S-wave pollution}
\medskip 

Another source of pollution is due to  events   from the S-wave decay $B\to K_0^*(\to K\pi)   \mu^+\mu^-$ where $K_0^*$ is a scalar resonance and its interference with the P-wave decay $B \to K^*(\to K\pi)\mu^+\mu^-$.
 In Ref.~\cite{1207.4004} a detailed and complete calculation of the S-wave background was performed and it was concluded that any observable will unavoidably suffer from its pollution. While this conclusion is correct in the case of uniangular distributions, it does not apply to full or folded distributions where the P- and the S-wave parts can be separated according to their different angular dependence.   As shown in Ref.~\cite{1209.1525}, S-wave pollution can easily be avoided for the $P_i^{(\prime)}$ observables if folded distributions are used instead of uniangular ones. A discussion of the experimental implications of the S-wave contribution was presented in Ref.~\cite{1210.5279} (see also Ref.~\cite{1307.0947}). A parametrization of these terms were given in \cite{1303.5794} together with a set of bounds on their size. Recently, it has been shown in \cite{1502.00920} using the symmetries of the distribution that only 4 (out of 6) of such parameters are  independent and  the explicit correlations among the over-complete six observables  have been provided.

\section{\bf $Z^\prime$ particle, other scenarios and the importance of considering all bins}
\medskip

In \cite{understanding} the existence of a $Z^\prime$ boson was proposed as a possible explanation of the large negative contribution to $C_9$, with couplings $\Delta_R^{sb} \sim 0$ (in the notation of \cite{defazio1,defazio2}) not to give contributions to $C_9^\prime$, with $\Delta _L^{sb}$ having the same phase as $V_{tb} V_{ts}^*$ to avoid large contributions to the $B_s$ mixing phase $\phi_s$, and a left-right symmetric coupling to muons $\Delta^{\mu}=\Delta_L^{\mu\mu}=\Delta_R^{\mu\mu}$ not to contribute to $C_{10}$. Finally, the  constraint coming from $\Delta m_s$  fixes the flavour-changing coupling $\Delta_L^{sb}$. A particle with couplings to muons $\Delta^{\mu}$ of order 0.1-0.2 and with a mass around 1-2 TeV would give the right contribution to $C_9$. Of course, this is an ad-hoc model and moreover, some of the coefficients that were taken to be zero in the first dataset may be switched on with more data. Later on there have been several attempts  to embed a $Z^\prime$ particle inside specific models (see \cite{zprime1,zprime2,zprime3,zprime4}).

Recently, using a large set of observables it was confirmed in \cite{straubnew} that a solution with $C_9^{\rm NP}<0$ gives the best fit. In their analysis  for the $B \to K^*\mu^+\mu^-$ mode the authors of \cite{straubnew} do not use  the long [1-6] bin which has a lower sensitivity to some Wilson coefficients but smaller bins (two out of the three possible bins).  Another interesting outcome of this work \cite{straubnew}, is the proposal of a second possible solution. There it is found that $C_9^{\rm NP}=-C_{10}^{\rm NP}$  could also provide  an explanation (with only slightly worse $\chi^2$ as compared to a fit where  only $C_9^{\rm NP}$ is switched on, see Table 2 in \cite{straubnew}) with all other coefficients switched off. This solution is an interesting possibility from a model-building point of view, but also because it would introduce NP in a cleaner coefficient $C_{10}$ (notice that this coefficient cannot receive any charm contribution). More in general, it has been found in \cite{grinstein}, under the assumption that the scale of NP is well above the electroweak scale,  that if one imposes that the operators of the effective lagrangian originate from  manifestly $SU(3)_c \times SU(2)_L \times U(1)_Y$ invariant high-scale operators, in the process of integrating the heavy degrees of freedom a set of relations between the coefficients of the effective operators arises at low scale. In particular, no tensor contribution are generated and scalar and pseudoscalar coefficients  are correlated. The remaining semileptonic coefficients are still independent in general, but functions of the coefficients of the high scale operators. If some of those coefficients of the dimension 6 operators vanish, the coefficients at the low-scale would become correlated.

We have performed an exploratory scan analysis of the scenario $C_9^{\rm NP}=-C_{10}^{\rm NP}$,  considering all three bins of the relevant observables $P_1$, $P_2$, $P_{4,5}^\prime$ together with ${\cal B}_{Bs \to \mu^+\mu^-}$, and  taking 2$\sigma$ for the theory errors and enlarging the experimental errors of the third bins to 2$\sigma$ to cover possible changes in the data  (see Section 3.4 and Fig. 1).  The result of this very preliminary analysis is:

\begin{itemize}
\item If only the first two bins of $B \to K^* \mu^+\mu^-$ for the relevant observables (and $B_s \to \mu^+\mu^-$) are included we find points that fulfill the constraints \begin{equation} \quad C_9^{\rm NP}=-C_{10}^{\rm NP}<0, \quad C_i^{\prime}=0, \end{equation} confirming the result in \cite{straubnew} using our method and observables. Notice that this only proves that  points exist that fulfill the constraints, but  we did not test the quality of the fit for this exploratory scan. Of course, one should check if those points survive constraints from other observables like $B \to K\mu^+\mu^-$ as it was done in \cite{straubnew}. \item On the other hand, if also 
 the third bins for the relevant observables are  included in the scan (not included in \cite{straubnew}),  all the previous points disappear. The reason is easy to understand: in the same way as a large $C_9^\prime >0$ substantially worsens the anomaly (3rd bin) in $P_5^\prime$, also a positive $C_{10}^{\rm NP}$ tends to enhance the anomaly. This suggests the need to switch one other coefficients like $C_9^\prime$, $C_{10}^\prime$.

\end{itemize}

 The new 3 fb$^{-1}$ data on $B \to K^* \mu^+\mu^-$ could easily confirm or erase and draw completely new scenarios. 



\section{Conclusions and Outlook}
\medskip

Using a basis of optimized observables LHCb released in 2013 the 1 fb$^{-1}$ data of $B \to K^*\mu^+\mu^-$ which exhibited a set of coherent tensions focused mainly in the observables $P_2$ and $P_5^\prime$.
The  analysis in \cite{understanding}  using all large-recoil and low-recoil bins pointed clearly to a large negative NP contribution to the Wilson coefficient $C_9$, and with less significance also to a negative contribution to $C_7$, while all other coefficients were consistent with the SM. Later on,  analyses done by other groups using different observables and methods \cite{1308.1501,1310.2478, 1310.3887, straubnew} and including the new 3 fb$^{-1}$ data of $B \to K \mu^+\mu^-$ and $B_s \to \phi \mu^+\mu^-$ confirmed this general pattern.

Here we have discussed additional subleading corrections that we included in the theoretical predictions given in \cite{1407.8526}  to confront the new data. Particular emphasis has been put on factorizable and non-factorizable (specially coming from charm loops) power corrections.  It is important to control these factorizable power corrections because they  affect all $P_i$ and all $S_i$ observables if computed from QCDF approach as done in \cite{1407.8526,jaeger}.  We have given details on our method to include the factorizable power corrections based on model-independent correlations and using an adequate scheme to define soft form factors. Moreover, we have explicitly shown using the example of the observable $S_5$  that the predictions obtained using two  different methods to deal with those corrections \cite{1407.8526} and \cite{straubnew} (including correlations in a specific form factor parametrization or using symmetries to include the dominant ones together with dimensional arguments)   are in excellent agreement. 

We suggest simple tests to be done on the new data to test an alternative explanation that was suggested to explain the $P_5^\prime$ anomaly by the presence of huge charm-loop contributions. The simplest one consists in a comparison between the  bins [4,6] and [6,8] of $P_5^\prime$. Moreover, the confirmation of the $R_K$ tension that could be explained simultaneously with the anomaly in $B \to K^* \mu^+\mu^-$ within the same NP scenario \cite{ghosh} (with couplings only to muons) but not via a large charm-loop contribution, would further challenge this alternative explanation.

We have also presented tests to perform on data, like comparing the measured value of $P_2$ with the one obtained from the measured values of $P_1$, $P_{4,5}^\prime$ that indicates that 
 future data should evolve such that the order of the  bins between $P_5^\prime$ and $P_2$ in the region above 4 GeV$^2$ should flip, with $P_2$ moving above $P_5^\prime$.

Finally, we have done an exploratory scan using only a subset of relevant observables to see if we could find solutions consistent with the scenario $C_9^{\rm NP}=-C_{10}^{\rm NP}$ and $C_9^\prime=C_{10}^\prime=0$ recently proposed 
 in \cite{straubnew}. While we confirmed that 
  this scenario  is perfectly viable if only the first two bins in $B \to K^*\mu^+\mu^-$ are taken into account (in agreement with \cite{straubnew}), we could not find solutions once the third bins are included. 
  
   We expect/hope that the new data can help  to find the next step in the path towards NP and start switching on with sufficient significance other Wilson coefficients besides $C_9^{\rm NP}$.

\section*{Acknowledgements}
J.V. is funded by the Deutsche Forschungsgemeinschaft (DFG) within research unit FOR 1873 (QFET). J.M. and L.H. acknowledge support from FPA2011-25948, 2014 SGR 1450.



\subsection*{{\bf Appendix}: SM predictions for $B\to K^*\mu^+\mu^-$ Observables}
\medskip


We provide here our  SM predictions obtained using the method described in \cite{1407.8526} for two different form factor parametrizations \cite{KMPW} and \cite{bz} including all theoretical errors: parametric errors, errors from form factors, factorizable power corrections, non-factorizable power corrections and charm-loop (only for KMPW) respectively. Notice that the error for the $P_i$ observables coming from form factors in BZ parametrization is below the permille level.\\



\newpage


\begin{tabular}{@{}crr@{}}
\hline\hline \\[-3mm] 
Observable  $\av{P_1}$ & KMPW - scheme 1 \hspace{10mm} & BZ - scheme 1  \hspace{10mm} \\[2mm] 
 \hline \\[-3mm] 
$[0.1,0.98] $  & $0.027_{-0.003}^{+0.005}{}_{-0.007}^{+0.006}{}_{-0.009}^{+0.009}{}_{-0.044}^{+0.035}{}_{-0.083}^{+0.061} $ & $0.036_{-0.004}^{+0.006}{}_{-0.000}^{+0.000}{}_{-0.009}^{+0.009}{}_{-0.046}^{+0.036}{}_{----}^{+---} $ \\[2mm] 
 \hline \\[-3mm] 
$[1.1,2] $  & $0.001_{-0.003}^{+0.003}{}_{-0.021}^{+0.016}{}_{-0.024}^{+0.020}{}_{-0.043}^{+0.033}{}_{-0.122}^{+0.091} $ & $0.030_{-0.004}^{+0.005}{}_{-0.000}^{+0.000}{}_{-0.020}^{+0.015}{}_{-0.043}^{+0.033}{}_{----}^{+---} $ \\[2mm] 
 \hline \\[-3mm] 
$[2,3] $  & $-0.008_{-0.002}^{+0.003}{}_{-0.013}^{+0.006}{}_{-0.025}^{+0.024}{}_{-0.025}^{+0.019}{}_{-0.096}^{+0.074} $ & $0.001_{-0.004}^{+0.003}{}_{-0.000}^{+0.000}{}_{-0.039}^{+0.031}{}_{-0.021}^{+0.016}{}_{----}^{+---} $ \\[2mm] 
 \hline \\[-3mm] 
$[3,4] $  & $0.005_{-0.004}^{+0.005}{}_{-0.015}^{+0.012}{}_{-0.051}^{+0.050}{}_{-0.006}^{+0.005}{}_{-0.028}^{+0.025} $ & $-0.034_{-0.003}^{+0.003}{}_{-0.000}^{+0.000}{}_{-0.067}^{+0.060}{}_{-0.003}^{+0.003}{}_{----}^{+---} $ \\[2mm] 
 \hline \\[-3mm] 
$[4,5] $  & $0.019_{-0.003}^{+0.005}{}_{-0.029}^{+0.033}{}_{-0.065}^{+0.062}{}_{-0.002}^{+0.002}{}_{-0.013}^{+0.021} $ & $-0.058_{-0.002}^{+0.001}{}_{-0.000}^{+0.000}{}_{-0.075}^{+0.075}{}_{-0.003}^{+0.003}{}_{----}^{+---} $ \\[2mm] 
 \hline \\[-3mm] 
$[5,6] $  & $0.023_{-0.002}^{+0.005}{}_{-0.036}^{+0.044}{}_{-0.071}^{+0.071}{}_{-0.003}^{+0.003}{}_{-0.035}^{+0.050} $ & $-0.077_{-0.002}^{+0.001}{}_{-0.000}^{+0.000}{}_{-0.078}^{+0.079}{}_{-0.004}^{+0.004}{}_{----}^{+---} $ \\[2mm] 
 \hline \\[-3mm] 
$[6,7] $  & $0.020_{-0.001}^{+0.005}{}_{-0.038}^{+0.050}{}_{-0.072}^{+0.074}{}_{-0.004}^{+0.004}{}_{-0.051}^{+0.072} $ & $-0.094_{-0.003}^{+0.002}{}_{-0.001}^{+0.000}{}_{-0.076}^{+0.078}{}_{-0.004}^{+0.004}{}_{----}^{+---} $ \\[2mm] 
 \hline \\[-3mm] 
$[7,8] $  & $0.012_{-0.001}^{+0.003}{}_{-0.039}^{+0.052}{}_{-0.070}^{+0.074}{}_{-0.006}^{+0.005}{}_{-0.070}^{+0.100} $ & $-0.113_{-0.004}^{+0.002}{}_{-0.001}^{+0.001}{}_{-0.074}^{+0.076}{}_{-0.006}^{+0.006}{}_{----}^{+---} $ \\[2mm] 
 \hline \\[-3mm] 
$[1.1,2.5] $  & $-0.002_{-0.002}^{+0.003}{}_{-0.020}^{+0.015}{}_{-0.021}^{+0.016}{}_{-0.039}^{+0.030}{}_{-0.119}^{+0.089} $ & $0.025_{-0.004}^{+0.004}{}_{-0.000}^{+0.000}{}_{-0.014}^{+0.010}{}_{-0.039}^{+0.030}{}_{----}^{+---} $ \\[2mm] 
 \hline \\[-3mm] 
$[2.5,4] $  & $0.001_{-0.003}^{+0.005}{}_{-0.010}^{+0.006}{}_{-0.045}^{+0.045}{}_{-0.010}^{+0.008}{}_{-0.045}^{+0.037} $ & $-0.026_{-0.004}^{+0.003}{}_{-0.000}^{+0.000}{}_{-0.062}^{+0.054}{}_{-0.006}^{+0.006}{}_{----}^{+---} $ \\[2mm] 
 \hline \\[-3mm] 
$[4,6] $  & $0.021_{-0.002}^{+0.005}{}_{-0.033}^{+0.039}{}_{-0.068}^{+0.066}{}_{-0.003}^{+0.003}{}_{-0.025}^{+0.037} $ & $-0.069_{-0.002}^{+0.001}{}_{-0.000}^{+0.000}{}_{-0.077}^{+0.077}{}_{-0.003}^{+0.004}{}_{----}^{+---} $ \\[2mm] 
 \hline \\[-3mm] 
$[6,8] $  & $0.016_{-0.001}^{+0.004}{}_{-0.039}^{+0.051}{}_{-0.071}^{+0.074}{}_{-0.005}^{+0.004}{}_{-0.062}^{+0.087} $ & $-0.104_{-0.004}^{+0.002}{}_{-0.001}^{+0.001}{}_{-0.075}^{+0.077}{}_{-0.005}^{+0.005}{}_{----}^{+---} $ \\[2mm] 
 \hline\hline 
\end{tabular}
\bigskip



\begin{tabular}{@{}crr@{}}
\hline\hline \\[-3mm] 
Observable $\av{P_2}$ & KMPW - scheme 1 \hspace{10mm} & BZ - scheme 1  \hspace{10mm} \\[2mm] 
 \hline \\[-3mm] 
$[0.1,0.98] $  & $0.118_{-0.006}^{+0.007}{}_{-0.005}^{+0.004}{}_{-0.012}^{+0.015}{}_{-0.002}^{+0.002}{}_{-0.003}^{+0.004} $ & $0.125_{-0.006}^{+0.007}{}_{-0.000}^{+0.000}{}_{-0.013}^{+0.014}{}_{-0.002}^{+0.002}{}_{----}^{+---} $ \\[2mm] 
 \hline \\[-3mm] 
$[1.1,2] $  & $0.428_{-0.017}^{+0.018}{}_{-0.013}^{+0.009}{}_{-0.032}^{+0.029}{}_{-0.005}^{+0.005}{}_{-0.011}^{+0.012} $ & $0.447_{-0.016}^{+0.015}{}_{-0.000}^{+0.000}{}_{-0.031}^{+0.020}{}_{-0.005}^{+0.004}{}_{----}^{+---} $ \\[2mm] 
 \hline \\[-3mm] 
$[2,3] $  & $0.438_{-0.029}^{+0.012}{}_{-0.018}^{+0.019}{}_{-0.067}^{+0.037}{}_{-0.008}^{+0.006}{}_{-0.034}^{+0.023} $ & $0.386_{-0.040}^{+0.022}{}_{-0.000}^{+0.000}{}_{-0.083}^{+0.063}{}_{-0.012}^{+0.009}{}_{----}^{+---} $ \\[2mm] 
 \hline \\[-3mm] 
$[3,4] $  & $0.153_{-0.061}^{+0.038}{}_{-0.038}^{+0.049}{}_{-0.107}^{+0.103}{}_{-0.015}^{+0.012}{}_{-0.077}^{+0.066} $ & $0.051_{-0.059}^{+0.039}{}_{-0.000}^{+0.000}{}_{-0.099}^{+0.115}{}_{-0.015}^{+0.012}{}_{----}^{+---} $ \\[2mm] 
 \hline \\[-3mm] 
$[4,5] $  & $-0.102_{-0.066}^{+0.036}{}_{-0.033}^{+0.047}{}_{-0.078}^{+0.092}{}_{-0.011}^{+0.009}{}_{-0.073}^{+0.070} $ & $-0.189_{-0.057}^{+0.030}{}_{-0.000}^{+0.000}{}_{-0.064}^{+0.086}{}_{-0.009}^{+0.007}{}_{----}^{+---} $ \\[2mm] 
 \hline \\[-3mm] 
$[5,6] $  & $-0.255_{-0.065}^{+0.027}{}_{-0.024}^{+0.036}{}_{-0.049}^{+0.064}{}_{-0.006}^{+0.006}{}_{-0.058}^{+0.061} $ & $-0.318_{-0.053}^{+0.021}{}_{-0.000}^{+0.000}{}_{-0.037}^{+0.054}{}_{-0.005}^{+0.004}{}_{----}^{+---} $ \\[2mm] 
 \hline \\[-3mm] 
$[6,7] $  & $-0.345_{-0.073}^{+0.025}{}_{-0.017}^{+0.026}{}_{-0.031}^{+0.043}{}_{-0.003}^{+0.004}{}_{-0.046}^{+0.053} $ & $-0.389_{-0.056}^{+0.019}{}_{-0.000}^{+0.000}{}_{-0.022}^{+0.032}{}_{-0.003}^{+0.003}{}_{----}^{+---} $ \\[2mm] 
 \hline \\[-3mm] 
$[7,8] $  & $-0.402_{-0.054}^{+0.027}{}_{-0.012}^{+0.019}{}_{-0.019}^{+0.028}{}_{-0.003}^{+0.004}{}_{-0.040}^{+0.049} $ & $-0.432_{-0.040}^{+0.020}{}_{-0.000}^{+0.000}{}_{-0.013}^{+0.019}{}_{-0.002}^{+0.003}{}_{----}^{+---} $ \\[2mm] 
 \hline \\[-3mm] 
$[1.1,2.5] $  & $0.443_{-0.013}^{+0.009}{}_{-0.007}^{+0.004}{}_{-0.021}^{+0.011}{}_{-0.003}^{+0.003}{}_{-0.006}^{+0.004} $ & $0.450_{-0.009}^{+0.004}{}_{-0.000}^{+0.000}{}_{-0.014}^{+0.002}{}_{-0.003}^{+0.002}{}_{----}^{+---} $ \\[2mm] 
 \hline \\[-3mm] 
$[2.5,4] $  & $0.227_{-0.058}^{+0.035}{}_{-0.036}^{+0.045}{}_{-0.105}^{+0.093}{}_{-0.015}^{+0.011}{}_{-0.071}^{+0.058} $ & $0.130_{-0.059}^{+0.039}{}_{-0.000}^{+0.000}{}_{-0.103}^{+0.111}{}_{-0.016}^{+0.012}{}_{----}^{+---} $ \\[2mm] 
 \hline \\[-3mm] 
$[4,6] $  & $-0.189_{-0.067}^{+0.031}{}_{-0.028}^{+0.041}{}_{-0.062}^{+0.077}{}_{-0.008}^{+0.007}{}_{-0.065}^{+0.066} $ & $-0.262_{-0.055}^{+0.025}{}_{-0.000}^{+0.000}{}_{-0.049}^{+0.068}{}_{-0.007}^{+0.005}{}_{----}^{+---} $ \\[2mm] 
 \hline \\[-3mm] 
$[6,8] $  & $-0.377_{-0.057}^{+0.026}{}_{-0.014}^{+0.022}{}_{-0.024}^{+0.034}{}_{-0.003}^{+0.004}{}_{-0.042}^{+0.051} $ & $-0.412_{-0.043}^{+0.020}{}_{-0.000}^{+0.000}{}_{-0.017}^{+0.025}{}_{-0.002}^{+0.003}{}_{----}^{+---} $ \\[2mm] 
 \hline\hline 
\end{tabular}
\bigskip



\begin{tabular}{@{}crr@{}}
\hline\hline \\[-3mm] 
Observable $\av{P_3}$ & KMPW - scheme 1 \hspace{10mm} & BZ - scheme 1  \hspace{10mm} \\[2mm] 
 \hline \\[-3mm] 
$[0.1,0.98] $  & $-0.001_{-0.001}^{+0.000}{}_{-0.000}^{+0.000}{}_{-0.001}^{+0.000}{}_{-0.019}^{+0.017}{}_{-0.003}^{+0.004} $ & $-0.001_{-0.001}^{+0.001}{}_{-0.000}^{+0.000}{}_{-0.001}^{+0.000}{}_{-0.020}^{+0.018}{}_{----}^{+---} $ \\[2mm] 
 \hline \\[-3mm] 
$[1.1,2] $  & $0.000_{-0.000}^{+0.000}{}_{-0.001}^{+0.001}{}_{-0.002}^{+0.002}{}_{-0.018}^{+0.016}{}_{-0.005}^{+0.008} $ & $-0.002_{-0.001}^{+0.001}{}_{-0.000}^{+0.000}{}_{-0.003}^{+0.002}{}_{-0.018}^{+0.017}{}_{----}^{+---} $ \\[2mm] 
 \hline \\[-3mm] 
$[2,3] $  & $0.002_{-0.001}^{+0.001}{}_{-0.002}^{+0.003}{}_{-0.004}^{+0.004}{}_{-0.010}^{+0.010}{}_{-0.007}^{+0.010} $ & $-0.002_{-0.002}^{+0.001}{}_{-0.000}^{+0.000}{}_{-0.004}^{+0.004}{}_{-0.009}^{+0.008}{}_{----}^{+---} $ \\[2mm] 
 \hline \\[-3mm] 
$[3,4] $  & $0.002_{-0.001}^{+0.002}{}_{-0.002}^{+0.003}{}_{-0.004}^{+0.004}{}_{-0.003}^{+0.003}{}_{-0.007}^{+0.009} $ & $-0.002_{-0.001}^{+0.001}{}_{-0.000}^{+0.000}{}_{-0.003}^{+0.004}{}_{-0.001}^{+0.001}{}_{----}^{+---} $ \\[2mm] 
 \hline \\[-3mm] 
$[4,5] $  & $0.002_{-0.001}^{+0.002}{}_{-0.001}^{+0.003}{}_{-0.003}^{+0.004}{}_{-0.001}^{+0.001}{}_{-0.005}^{+0.007} $ & $-0.001_{-0.001}^{+0.001}{}_{-0.000}^{+0.000}{}_{-0.003}^{+0.003}{}_{-0.002}^{+0.002}{}_{----}^{+---} $ \\[2mm] 
 \hline \\[-3mm] 
$[5,6] $  & $0.002_{-0.001}^{+0.003}{}_{-0.001}^{+0.002}{}_{-0.003}^{+0.003}{}_{-0.002}^{+0.002}{}_{-0.005}^{+0.006} $ & $-0.001_{-0.001}^{+0.001}{}_{-0.000}^{+0.000}{}_{-0.002}^{+0.003}{}_{-0.002}^{+0.002}{}_{----}^{+---} $ \\[2mm] 
 \hline \\[-3mm] 
$[6,7] $  & $0.002_{-0.002}^{+0.004}{}_{-0.001}^{+0.002}{}_{-0.002}^{+0.003}{}_{-0.002}^{+0.002}{}_{-0.005}^{+0.006} $ & $-0.001_{-0.002}^{+0.001}{}_{-0.000}^{+0.000}{}_{-0.002}^{+0.002}{}_{-0.002}^{+0.002}{}_{----}^{+---} $ \\[2mm] 
 \hline \\[-3mm] 
$[7,8] $  & $0.001_{-0.007}^{+0.003}{}_{-0.001}^{+0.001}{}_{-0.002}^{+0.003}{}_{-0.003}^{+0.004}{}_{-0.006}^{+0.007} $ & $-0.001_{-0.002}^{+0.003}{}_{-0.000}^{+0.000}{}_{-0.002}^{+0.002}{}_{-0.003}^{+0.004}{}_{----}^{+---} $ \\[2mm] 
 \hline \\[-3mm] 
$[1.1,2.5] $  & $0.000_{-0.000}^{+0.000}{}_{-0.001}^{+0.002}{}_{-0.003}^{+0.003}{}_{-0.016}^{+0.015}{}_{-0.006}^{+0.008} $ & $-0.002_{-0.001}^{+0.001}{}_{-0.000}^{+0.000}{}_{-0.003}^{+0.003}{}_{-0.016}^{+0.015}{}_{----}^{+---} $ \\[2mm] 
 \hline \\[-3mm] 
$[2.5,4] $  & $0.002_{-0.001}^{+0.002}{}_{-0.002}^{+0.003}{}_{-0.004}^{+0.004}{}_{-0.004}^{+0.004}{}_{-0.007}^{+0.009} $ & $-0.002_{-0.001}^{+0.001}{}_{-0.000}^{+0.000}{}_{-0.003}^{+0.004}{}_{-0.003}^{+0.003}{}_{----}^{+---} $ \\[2mm] 
 \hline \\[-3mm] 
$[4,6] $  & $0.002_{-0.001}^{+0.003}{}_{-0.001}^{+0.002}{}_{-0.003}^{+0.003}{}_{-0.001}^{+0.002}{}_{-0.005}^{+0.007} $ & $-0.001_{-0.001}^{+0.001}{}_{-0.000}^{+0.000}{}_{-0.002}^{+0.003}{}_{-0.002}^{+0.002}{}_{----}^{+---} $ \\[2mm] 
 \hline \\[-3mm] 
$[6,8] $  & $0.001_{-0.005}^{+0.002}{}_{-0.001}^{+0.002}{}_{-0.002}^{+0.003}{}_{-0.002}^{+0.003}{}_{-0.005}^{+0.007} $ & $-0.001_{-0.001}^{+0.002}{}_{-0.000}^{+0.000}{}_{-0.002}^{+0.002}{}_{-0.002}^{+0.003}{}_{----}^{+---} $ \\[2mm] 
 \hline\hline 
\end{tabular}
\bigskip



\begin{tabular}{@{}crr@{}}
\hline\hline \\[-3mm] 
Observable $\av{P'_4}$ & KMPW - scheme 1 \hspace{10mm} & BZ - scheme 1  \hspace{10mm} \\[2mm] 
 \hline \\[-3mm] 
$[0.1,0.98] $  & $-0.495_{-0.016}^{+0.013}{}_{-0.017}^{+0.029}{}_{-0.023}^{+0.022}{}_{-0.010}^{+0.010}{}_{-0.211}^{+0.335} $ & $-0.476_{-0.017}^{+0.016}{}_{-0.001}^{+0.001}{}_{-0.022}^{+0.020}{}_{-0.011}^{+0.010}{}_{----}^{+---} $ \\[2mm] 
 \hline \\[-3mm] 
$[1.1,2] $  & $-0.162_{-0.025}^{+0.036}{}_{-0.050}^{+0.055}{}_{-0.073}^{+0.096}{}_{-0.011}^{+0.011}{}_{-0.107}^{+0.117} $ & $-0.076_{-0.030}^{+0.046}{}_{-0.002}^{+0.002}{}_{-0.086}^{+0.101}{}_{-0.012}^{+0.012}{}_{----}^{+---} $ \\[2mm] 
 \hline \\[-3mm] 
$[2,3] $  & $0.271_{-0.043}^{+0.052}{}_{-0.082}^{+0.075}{}_{-0.126}^{+0.144}{}_{-0.011}^{+0.012}{}_{-0.097}^{+0.110} $ & $0.417_{-0.045}^{+0.051}{}_{-0.002}^{+0.002}{}_{-0.143}^{+0.135}{}_{-0.012}^{+0.014}{}_{----}^{+---} $ \\[2mm] 
 \hline \\[-3mm] 
$[3,4] $  & $0.611_{-0.036}^{+0.042}{}_{-0.083}^{+0.068}{}_{-0.120}^{+0.117}{}_{-0.008}^{+0.009}{}_{-0.087}^{+0.089} $ & $0.753_{-0.030}^{+0.034}{}_{-0.001}^{+0.001}{}_{-0.119}^{+0.096}{}_{-0.008}^{+0.008}{}_{----}^{+---} $ \\[2mm] 
 \hline \\[-3mm] 
$[4,5] $  & $0.785_{-0.022}^{+0.026}{}_{-0.071}^{+0.054}{}_{-0.092}^{+0.082}{}_{-0.005}^{+0.005}{}_{-0.078}^{+0.072} $ & $0.905_{-0.016}^{+0.018}{}_{-0.001}^{+0.001}{}_{-0.083}^{+0.065}{}_{-0.004}^{+0.005}{}_{----}^{+---} $ \\[2mm] 
 \hline \\[-3mm] 
$[5,6] $  & $0.871_{-0.013}^{+0.016}{}_{-0.060}^{+0.044}{}_{-0.070}^{+0.061}{}_{-0.004}^{+0.004}{}_{-0.077}^{+0.061} $ & $0.974_{-0.009}^{+0.011}{}_{-0.000}^{+0.000}{}_{-0.061}^{+0.051}{}_{-0.003}^{+0.004}{}_{----}^{+---} $ \\[2mm] 
 \hline \\[-3mm] 
$[6,7] $  & $0.918_{-0.017}^{+0.013}{}_{-0.052}^{+0.037}{}_{-0.056}^{+0.050}{}_{-0.004}^{+0.004}{}_{-0.079}^{+0.055} $ & $1.011_{-0.014}^{+0.009}{}_{-0.000}^{+0.000}{}_{-0.048}^{+0.043}{}_{-0.003}^{+0.004}{}_{----}^{+---} $ \\[2mm] 
 \hline \\[-3mm] 
$[7,8] $  & $0.949_{-0.017}^{+0.012}{}_{-0.046}^{+0.032}{}_{-0.049}^{+0.043}{}_{-0.005}^{+0.006}{}_{-0.091}^{+0.057} $ & $1.035_{-0.016}^{+0.009}{}_{-0.000}^{+0.000}{}_{-0.042}^{+0.038}{}_{-0.004}^{+0.005}{}_{----}^{+---} $ \\[2mm] 
 \hline \\[-3mm] 
$[1.1,2.5] $  & $-0.058_{-0.030}^{+0.039}{}_{-0.059}^{+0.064}{}_{-0.088}^{+0.113}{}_{-0.011}^{+0.011}{}_{-0.102}^{+0.117} $ & $0.043_{-0.036}^{+0.050}{}_{-0.002}^{+0.002}{}_{-0.104}^{+0.116}{}_{-0.012}^{+0.013}{}_{----}^{+---} $ \\[2mm] 
 \hline \\[-3mm] 
$[2.5,4] $  & $0.537_{-0.039}^{+0.047}{}_{-0.085}^{+0.073}{}_{-0.125}^{+0.127}{}_{-0.009}^{+0.010}{}_{-0.092}^{+0.094} $ & $0.683_{-0.035}^{+0.040}{}_{-0.002}^{+0.001}{}_{-0.129}^{+0.107}{}_{-0.009}^{+0.010}{}_{----}^{+---} $ \\[2mm] 
 \hline \\[-3mm] 
$[4,6] $  & $0.830_{-0.017}^{+0.020}{}_{-0.065}^{+0.050}{}_{-0.080}^{+0.071}{}_{-0.004}^{+0.005}{}_{-0.077}^{+0.066} $ & $0.940_{-0.012}^{+0.013}{}_{-0.001}^{+0.001}{}_{-0.071}^{+0.057}{}_{-0.004}^{+0.004}{}_{----}^{+---} $ \\[2mm] 
 \hline \\[-3mm] 
$[6,8] $  & $0.933_{-0.011}^{+0.013}{}_{-0.048}^{+0.035}{}_{-0.052}^{+0.046}{}_{-0.004}^{+0.005}{}_{-0.086}^{+0.055} $ & $1.022_{-0.010}^{+0.009}{}_{-0.000}^{+0.000}{}_{-0.045}^{+0.040}{}_{-0.003}^{+0.005}{}_{----}^{+---} $ \\[2mm] 
 \hline\hline 
\end{tabular}
\bigskip



\begin{tabular}{@{}crr@{}}
\hline\hline \\[-3mm] 
Observable $\av{P'_5}$ & KMPW - scheme 1 \hspace{10mm} & BZ - scheme 1  \hspace{10mm} \\[2mm] 
 \hline \\[-3mm] 
$[0.1,0.98] $  & $0.675_{-0.016}^{+0.011}{}_{-0.013}^{+0.014}{}_{-0.037}^{+0.035}{}_{-0.013}^{+0.012}{}_{-0.187}^{+0.152} $ & $0.679_{-0.017}^{+0.011}{}_{-0.001}^{+0.001}{}_{-0.034}^{+0.030}{}_{-0.015}^{+0.013}{}_{----}^{+---} $ \\[2mm] 
 \hline \\[-3mm] 
$[1.1,2] $  & $0.308_{-0.048}^{+0.032}{}_{-0.006}^{+0.013}{}_{-0.084}^{+0.078}{}_{-0.018}^{+0.016}{}_{-0.126}^{+0.092} $ & $0.289_{-0.051}^{+0.034}{}_{-0.002}^{+0.002}{}_{-0.088}^{+0.076}{}_{-0.019}^{+0.017}{}_{----}^{+---} $ \\[2mm] 
 \hline \\[-3mm] 
$[2,3] $  & $-0.164_{-0.076}^{+0.053}{}_{-0.012}^{+0.011}{}_{-0.113}^{+0.114}{}_{-0.023}^{+0.019}{}_{-0.126}^{+0.103} $ & $-0.204_{-0.075}^{+0.054}{}_{-0.002}^{+0.002}{}_{-0.112}^{+0.107}{}_{-0.024}^{+0.019}{}_{----}^{+---} $ \\[2mm] 
 \hline \\[-3mm] 
$[3,4] $  & $-0.553_{-0.072}^{+0.047}{}_{-0.005}^{+0.011}{}_{-0.094}^{+0.103}{}_{-0.018}^{+0.014}{}_{-0.109}^{+0.095} $ & $-0.570_{-0.066}^{+0.043}{}_{-0.002}^{+0.001}{}_{-0.088}^{+0.085}{}_{-0.016}^{+0.014}{}_{----}^{+---} $ \\[2mm] 
 \hline \\[-3mm] 
$[4,5] $  & $-0.761_{-0.064}^{+0.035}{}_{-0.012}^{+0.016}{}_{-0.069}^{+0.073}{}_{-0.011}^{+0.009}{}_{-0.089}^{+0.077} $ & $-0.746_{-0.056}^{+0.030}{}_{-0.001}^{+0.001}{}_{-0.068}^{+0.068}{}_{-0.009}^{+0.008}{}_{----}^{+---} $ \\[2mm] 
 \hline \\[-3mm] 
$[5,6] $  & $-0.865_{-0.058}^{+0.025}{}_{-0.020}^{+0.019}{}_{-0.052}^{+0.054}{}_{-0.007}^{+0.006}{}_{-0.077}^{+0.063} $ & $-0.828_{-0.050}^{+0.022}{}_{-0.001}^{+0.001}{}_{-0.056}^{+0.057}{}_{-0.007}^{+0.006}{}_{----}^{+---} $ \\[2mm] 
 \hline \\[-3mm] 
$[6,7] $  & $-0.919_{-0.062}^{+0.023}{}_{-0.024}^{+0.021}{}_{-0.041}^{+0.043}{}_{-0.006}^{+0.005}{}_{-0.076}^{+0.056} $ & $-0.870_{-0.053}^{+0.020}{}_{-0.000}^{+0.000}{}_{-0.048}^{+0.049}{}_{-0.005}^{+0.005}{}_{----}^{+---} $ \\[2mm] 
 \hline \\[-3mm] 
$[7,8] $  & $-0.951_{-0.044}^{+0.025}{}_{-0.026}^{+0.021}{}_{-0.037}^{+0.037}{}_{-0.006}^{+0.006}{}_{-0.082}^{+0.062} $ & $-0.893_{-0.037}^{+0.022}{}_{-0.000}^{+0.000}{}_{-0.042}^{+0.043}{}_{-0.005}^{+0.006}{}_{----}^{+---} $ \\[2mm] 
 \hline \\[-3mm] 
$[1.1,2.5] $  & $0.196_{-0.057}^{+0.038}{}_{-0.010}^{+0.010}{}_{-0.094}^{+0.088}{}_{-0.019}^{+0.017}{}_{-0.125}^{+0.094} $ & $0.171_{-0.060}^{+0.041}{}_{-0.002}^{+0.002}{}_{-0.098}^{+0.087}{}_{-0.021}^{+0.018}{}_{----}^{+---} $ \\[2mm] 
 \hline \\[-3mm] 
$[2.5,4] $  & $-0.467_{-0.073}^{+0.050}{}_{-0.008}^{+0.009}{}_{-0.100}^{+0.109}{}_{-0.020}^{+0.016}{}_{-0.114}^{+0.099} $ & $-0.491_{-0.068}^{+0.047}{}_{-0.002}^{+0.002}{}_{-0.095}^{+0.093}{}_{-0.019}^{+0.015}{}_{----}^{+---} $ \\[2mm] 
 \hline \\[-3mm] 
$[4,6] $  & $-0.816_{-0.061}^{+0.029}{}_{-0.017}^{+0.017}{}_{-0.060}^{+0.061}{}_{-0.008}^{+0.007}{}_{-0.082}^{+0.069} $ & $-0.788_{-0.053}^{+0.025}{}_{-0.001}^{+0.001}{}_{-0.061}^{+0.062}{}_{-0.008}^{+0.007}{}_{----}^{+---} $ \\[2mm] 
 \hline \\[-3mm] 
$[6,8] $  & $-0.935_{-0.048}^{+0.024}{}_{-0.025}^{+0.021}{}_{-0.039}^{+0.039}{}_{-0.006}^{+0.006}{}_{-0.079}^{+0.059} $ & $-0.881_{-0.040}^{+0.021}{}_{-0.000}^{+0.000}{}_{-0.045}^{+0.046}{}_{-0.005}^{+0.006}{}_{----}^{+---} $ \\[2mm] 
 \hline\hline 
\end{tabular}
\bigskip



\begin{tabular}{@{}crr@{}}
\hline\hline \\[-3mm] 
Observable $\av{P'_6}$ & KMPW - scheme 1 \hspace{10mm} & BZ - scheme 1  \hspace{10mm} \\[2mm] 
 \hline \\[-3mm] 
$[0.1,0.98] $  & $-0.058_{-0.031}^{+0.022}{}_{-0.008}^{+0.007}{}_{-0.006}^{+0.005}{}_{-0.017}^{+0.016}{}_{-0.015}^{+0.015} $ & $-0.060_{-0.031}^{+0.024}{}_{-0.001}^{+0.001}{}_{-0.006}^{+0.005}{}_{-0.018}^{+0.017}{}_{----}^{+---} $ \\[2mm] 
 \hline \\[-3mm] 
$[1.1,2] $  & $-0.072_{-0.037}^{+0.027}{}_{-0.010}^{+0.009}{}_{-0.008}^{+0.006}{}_{-0.016}^{+0.016}{}_{-0.012}^{+0.009} $ & $-0.076_{-0.038}^{+0.030}{}_{-0.001}^{+0.001}{}_{-0.008}^{+0.007}{}_{-0.017}^{+0.017}{}_{----}^{+---} $ \\[2mm] 
 \hline \\[-3mm] 
$[2,3] $  & $-0.070_{-0.038}^{+0.029}{}_{-0.011}^{+0.009}{}_{-0.004}^{+0.004}{}_{-0.012}^{+0.013}{}_{-0.009}^{+0.007} $ & $-0.073_{-0.039}^{+0.031}{}_{-0.001}^{+0.001}{}_{-0.004}^{+0.004}{}_{-0.013}^{+0.013}{}_{----}^{+---} $ \\[2mm] 
 \hline \\[-3mm] 
$[3,4] $  & $-0.056_{-0.035}^{+0.027}{}_{-0.011}^{+0.009}{}_{-0.002}^{+0.003}{}_{-0.008}^{+0.008}{}_{-0.005}^{+0.005} $ & $-0.056_{-0.034}^{+0.028}{}_{-0.001}^{+0.001}{}_{-0.003}^{+0.003}{}_{-0.007}^{+0.008}{}_{----}^{+---} $ \\[2mm] 
 \hline \\[-3mm] 
$[4,5] $  & $-0.041_{-0.034}^{+0.024}{}_{-0.010}^{+0.007}{}_{-0.003}^{+0.003}{}_{-0.005}^{+0.007}{}_{-0.003}^{+0.004} $ & $-0.040_{-0.031}^{+0.024}{}_{-0.001}^{+0.001}{}_{-0.004}^{+0.004}{}_{-0.006}^{+0.007}{}_{----}^{+---} $ \\[2mm] 
 \hline \\[-3mm] 
$[5,6] $  & $-0.030_{-0.042}^{+0.022}{}_{-0.009}^{+0.007}{}_{-0.004}^{+0.003}{}_{-0.009}^{+0.008}{}_{-0.003}^{+0.003} $ & $-0.029_{-0.040}^{+0.022}{}_{-0.001}^{+0.001}{}_{-0.004}^{+0.003}{}_{-0.009}^{+0.008}{}_{----}^{+---} $ \\[2mm] 
 \hline \\[-3mm] 
$[6,7] $  & $-0.022_{-0.075}^{+0.041}{}_{-0.008}^{+0.006}{}_{-0.003}^{+0.003}{}_{-0.012}^{+0.011}{}_{-0.004}^{+0.004} $ & $-0.020_{-0.067}^{+0.039}{}_{-0.001}^{+0.000}{}_{-0.004}^{+0.003}{}_{-0.013}^{+0.011}{}_{----}^{+---} $ \\[2mm] 
 \hline \\[-3mm] 
$[7,8] $  & $-0.015_{-0.070}^{+0.095}{}_{-0.007}^{+0.005}{}_{-0.004}^{+0.003}{}_{-0.017}^{+0.014}{}_{-0.005}^{+0.005} $ & $-0.014_{-0.069}^{+0.090}{}_{-0.000}^{+0.000}{}_{-0.004}^{+0.003}{}_{-0.017}^{+0.015}{}_{----}^{+---} $ \\[2mm] 
 \hline \\[-3mm] 
$[1.1,2.5] $  & $-0.072_{-0.037}^{+0.027}{}_{-0.010}^{+0.009}{}_{-0.007}^{+0.006}{}_{-0.015}^{+0.015}{}_{-0.011}^{+0.008} $ & $-0.076_{-0.038}^{+0.030}{}_{-0.001}^{+0.001}{}_{-0.007}^{+0.007}{}_{-0.016}^{+0.016}{}_{----}^{+---} $ \\[2mm] 
 \hline \\[-3mm] 
$[2.5,4] $  & $-0.060_{-0.036}^{+0.028}{}_{-0.011}^{+0.009}{}_{-0.002}^{+0.002}{}_{-0.009}^{+0.009}{}_{-0.006}^{+0.005} $ & $-0.060_{-0.035}^{+0.029}{}_{-0.001}^{+0.001}{}_{-0.002}^{+0.003}{}_{-0.008}^{+0.009}{}_{----}^{+---} $ \\[2mm] 
 \hline \\[-3mm] 
$[4,6] $  & $-0.035_{-0.039}^{+0.023}{}_{-0.009}^{+0.007}{}_{-0.004}^{+0.003}{}_{-0.007}^{+0.007}{}_{-0.003}^{+0.003} $ & $-0.034_{-0.036}^{+0.022}{}_{-0.001}^{+0.001}{}_{-0.004}^{+0.004}{}_{-0.008}^{+0.007}{}_{----}^{+---} $ \\[2mm] 
 \hline \\[-3mm] 
$[6,8] $  & $-0.018_{-0.046}^{+0.056}{}_{-0.007}^{+0.006}{}_{-0.003}^{+0.003}{}_{-0.014}^{+0.013}{}_{-0.004}^{+0.004} $ & $-0.017_{-0.045}^{+0.053}{}_{-0.000}^{+0.000}{}_{-0.004}^{+0.003}{}_{-0.015}^{+0.013}{}_{----}^{+---} $ \\[2mm] 
 \hline\hline 
\end{tabular}
\bigskip



\begin{tabular}{@{}crr@{}}
\hline\hline \\[-3mm] 
Observable $\av{P'_8}$ & KMPW - scheme 1 \hspace{10mm} & BZ - scheme 1  \hspace{10mm} \\[2mm] 
 \hline \\[-3mm] 
$[0.1,0.98] $  & $0.020_{-0.017}^{+0.027}{}_{-0.008}^{+0.009}{}_{-0.007}^{+0.008}{}_{-0.017}^{+0.016}{}_{-0.030}^{+0.021} $ & $0.023_{-0.017}^{+0.026}{}_{-0.001}^{+0.001}{}_{-0.006}^{+0.007}{}_{-0.017}^{+0.016}{}_{----}^{+---} $ \\[2mm] 
 \hline \\[-3mm] 
$[1.1,2] $  & $0.041_{-0.022}^{+0.035}{}_{-0.007}^{+0.009}{}_{-0.009}^{+0.010}{}_{-0.018}^{+0.016}{}_{-0.009}^{+0.010} $ & $0.044_{-0.024}^{+0.035}{}_{-0.001}^{+0.001}{}_{-0.008}^{+0.009}{}_{-0.018}^{+0.016}{}_{----}^{+---} $ \\[2mm] 
 \hline \\[-3mm] 
$[2,3] $  & $0.048_{-0.024}^{+0.038}{}_{-0.008}^{+0.009}{}_{-0.007}^{+0.007}{}_{-0.015}^{+0.014}{}_{-0.007}^{+0.007} $ & $0.049_{-0.025}^{+0.037}{}_{-0.001}^{+0.001}{}_{-0.006}^{+0.007}{}_{-0.015}^{+0.014}{}_{----}^{+---} $ \\[2mm] 
 \hline \\[-3mm] 
$[3,4] $  & $0.042_{-0.023}^{+0.036}{}_{-0.007}^{+0.009}{}_{-0.004}^{+0.004}{}_{-0.011}^{+0.010}{}_{-0.005}^{+0.005} $ & $0.041_{-0.023}^{+0.033}{}_{-0.001}^{+0.001}{}_{-0.004}^{+0.004}{}_{-0.010}^{+0.010}{}_{----}^{+---} $ \\[2mm] 
 \hline \\[-3mm] 
$[4,5] $  & $0.033_{-0.020}^{+0.034}{}_{-0.006}^{+0.008}{}_{-0.003}^{+0.003}{}_{-0.008}^{+0.007}{}_{-0.003}^{+0.003} $ & $0.032_{-0.019}^{+0.031}{}_{-0.000}^{+0.001}{}_{-0.002}^{+0.003}{}_{-0.008}^{+0.007}{}_{----}^{+---} $ \\[2mm] 
 \hline \\[-3mm] 
$[5,6] $  & $0.026_{-0.019}^{+0.039}{}_{-0.005}^{+0.007}{}_{-0.002}^{+0.002}{}_{-0.007}^{+0.006}{}_{-0.003}^{+0.003} $ & $0.024_{-0.018}^{+0.037}{}_{-0.000}^{+0.000}{}_{-0.002}^{+0.002}{}_{-0.007}^{+0.006}{}_{----}^{+---} $ \\[2mm] 
 \hline \\[-3mm] 
$[6,7] $  & $0.020_{-0.036}^{+0.063}{}_{-0.004}^{+0.006}{}_{-0.002}^{+0.002}{}_{-0.008}^{+0.006}{}_{-0.003}^{+0.003} $ & $0.019_{-0.033}^{+0.056}{}_{-0.000}^{+0.000}{}_{-0.002}^{+0.002}{}_{-0.007}^{+0.005}{}_{----}^{+---} $ \\[2mm] 
 \hline \\[-3mm] 
$[7,8] $  & $0.015_{-0.090}^{+0.058}{}_{-0.004}^{+0.005}{}_{-0.002}^{+0.002}{}_{-0.009}^{+0.009}{}_{-0.004}^{+0.003} $ & $0.014_{-0.082}^{+0.056}{}_{-0.000}^{+0.000}{}_{-0.002}^{+0.002}{}_{-0.009}^{+0.008}{}_{----}^{+---} $ \\[2mm] 
 \hline \\[-3mm] 
$[1.1,2.5] $  & $0.043_{-0.023}^{+0.036}{}_{-0.007}^{+0.009}{}_{-0.008}^{+0.009}{}_{-0.017}^{+0.015}{}_{-0.009}^{+0.009} $ & $0.045_{-0.024}^{+0.036}{}_{-0.001}^{+0.001}{}_{-0.008}^{+0.009}{}_{-0.018}^{+0.015}{}_{----}^{+---} $ \\[2mm] 
 \hline \\[-3mm] 
$[2.5,4] $  & $0.044_{-0.023}^{+0.036}{}_{-0.008}^{+0.009}{}_{-0.005}^{+0.005}{}_{-0.012}^{+0.011}{}_{-0.005}^{+0.005} $ & $0.043_{-0.023}^{+0.034}{}_{-0.001}^{+0.001}{}_{-0.004}^{+0.005}{}_{-0.012}^{+0.011}{}_{----}^{+---} $ \\[2mm] 
 \hline \\[-3mm] 
$[4,6] $  & $0.029_{-0.019}^{+0.037}{}_{-0.005}^{+0.007}{}_{-0.002}^{+0.002}{}_{-0.008}^{+0.006}{}_{-0.003}^{+0.003} $ & $0.028_{-0.019}^{+0.034}{}_{-0.000}^{+0.000}{}_{-0.002}^{+0.002}{}_{-0.007}^{+0.006}{}_{----}^{+---} $ \\[2mm] 
 \hline \\[-3mm] 
$[6,8] $  & $0.017_{-0.059}^{+0.039}{}_{-0.004}^{+0.006}{}_{-0.002}^{+0.002}{}_{-0.009}^{+0.007}{}_{-0.004}^{+0.003} $ & $0.016_{-0.054}^{+0.037}{}_{-0.000}^{+0.000}{}_{-0.002}^{+0.002}{}_{-0.008}^{+0.007}{}_{----}^{+---} $ \\[2mm] 
 \hline\hline 
\end{tabular}
\bigskip



\begin{tabular}{@{}crr@{}}
\hline\hline \\[-3mm] 
Observable $\av{A_\text{FB}}$ & KMPW - scheme 1 \hspace{10mm} & BZ - scheme 1  \hspace{10mm} \\[2mm] 
 \hline \\[-3mm] 
$[0.1,0.98] $  & $-0.099_{-0.002}^{+0.002}{}_{-0.026}^{+0.036}{}_{-0.006}^{+0.006}{}_{-0.001}^{+0.001}{}_{-0.009}^{+0.014} $ & $-0.097_{-0.002}^{+0.003}{}_{-0.004}^{+0.004}{}_{-0.004}^{+0.005}{}_{-0.000}^{+0.001}{}_{----}^{+---} $ \\[2mm] 
 \hline \\[-3mm] 
$[1.1,2] $  & $-0.199_{-0.007}^{+0.012}{}_{-0.182}^{+0.123}{}_{-0.020}^{+0.022}{}_{-0.003}^{+0.004}{}_{-0.021}^{+0.023} $ & $-0.172_{-0.010}^{+0.013}{}_{-0.015}^{+0.016}{}_{-0.019}^{+0.021}{}_{-0.003}^{+0.005}{}_{----}^{+---} $ \\[2mm] 
 \hline \\[-3mm] 
$[2,3] $  & $-0.140_{-0.013}^{+0.019}{}_{-0.152}^{+0.090}{}_{-0.030}^{+0.031}{}_{-0.005}^{+0.006}{}_{-0.022}^{+0.018} $ & $-0.106_{-0.014}^{+0.017}{}_{-0.010}^{+0.010}{}_{-0.030}^{+0.029}{}_{-0.004}^{+0.006}{}_{----}^{+---} $ \\[2mm] 
 \hline \\[-3mm] 
$[3,4] $  & $-0.048_{-0.014}^{+0.021}{}_{-0.032}^{+0.027}{}_{-0.035}^{+0.034}{}_{-0.004}^{+0.005}{}_{-0.023}^{+0.024} $ & $-0.015_{-0.012}^{+0.018}{}_{-0.001}^{+0.001}{}_{-0.035}^{+0.030}{}_{-0.004}^{+0.005}{}_{----}^{+---} $ \\[2mm] 
 \hline \\[-3mm] 
$[4,5] $  & $0.038_{-0.013}^{+0.023}{}_{-0.034}^{+0.066}{}_{-0.034}^{+0.031}{}_{-0.003}^{+0.004}{}_{-0.027}^{+0.030} $ & $0.070_{-0.011}^{+0.019}{}_{-0.006}^{+0.006}{}_{-0.033}^{+0.027}{}_{-0.003}^{+0.003}{}_{----}^{+---} $ \\[2mm] 
 \hline \\[-3mm] 
$[5,6] $  & $0.111_{-0.011}^{+0.026}{}_{-0.086}^{+0.126}{}_{-0.030}^{+0.027}{}_{-0.002}^{+0.002}{}_{-0.032}^{+0.037} $ & $0.142_{-0.010}^{+0.021}{}_{-0.011}^{+0.010}{}_{-0.030}^{+0.023}{}_{-0.002}^{+0.002}{}_{----}^{+---} $ \\[2mm] 
 \hline \\[-3mm] 
$[6,7] $  & $0.174_{-0.012}^{+0.034}{}_{-0.130}^{+0.159}{}_{-0.026}^{+0.023}{}_{-0.002}^{+0.002}{}_{-0.038}^{+0.044} $ & $0.204_{-0.010}^{+0.026}{}_{-0.014}^{+0.013}{}_{-0.025}^{+0.019}{}_{-0.002}^{+0.002}{}_{----}^{+---} $ \\[2mm] 
 \hline \\[-3mm] 
$[7,8] $  & $0.228_{-0.014}^{+0.028}{}_{-0.166}^{+0.176}{}_{-0.022}^{+0.019}{}_{-0.003}^{+0.004}{}_{-0.046}^{+0.057} $ & $0.256_{-0.012}^{+0.023}{}_{-0.016}^{+0.015}{}_{-0.021}^{+0.015}{}_{-0.003}^{+0.004}{}_{----}^{+---} $ \\[2mm] 
 \hline \\[-3mm] 
$[1.1,2.5] $  & $-0.187_{-0.009}^{+0.014}{}_{-0.182}^{+0.117}{}_{-0.023}^{+0.024}{}_{-0.004}^{+0.005}{}_{-0.021}^{+0.022} $ & $-0.157_{-0.012}^{+0.015}{}_{-0.014}^{+0.015}{}_{-0.021}^{+0.022}{}_{-0.004}^{+0.005}{}_{----}^{+---} $ \\[2mm] 
 \hline \\[-3mm] 
$[2.5,4] $  & $-0.071_{-0.014}^{+0.020}{}_{-0.061}^{+0.043}{}_{-0.035}^{+0.033}{}_{-0.004}^{+0.006}{}_{-0.021}^{+0.022} $ & $-0.038_{-0.013}^{+0.018}{}_{-0.003}^{+0.003}{}_{-0.034}^{+0.030}{}_{-0.004}^{+0.005}{}_{----}^{+---} $ \\[2mm] 
 \hline \\[-3mm] 
$[4,6] $  & $0.076_{-0.012}^{+0.025}{}_{-0.061}^{+0.100}{}_{-0.032}^{+0.029}{}_{-0.002}^{+0.003}{}_{-0.030}^{+0.034} $ & $0.107_{-0.010}^{+0.020}{}_{-0.009}^{+0.008}{}_{-0.032}^{+0.025}{}_{-0.002}^{+0.002}{}_{----}^{+---} $ \\[2mm] 
 \hline \\[-3mm] 
$[6,8] $  & $0.202_{-0.013}^{+0.028}{}_{-0.149}^{+0.169}{}_{-0.024}^{+0.021}{}_{-0.003}^{+0.003}{}_{-0.042}^{+0.051} $ & $0.231_{-0.012}^{+0.023}{}_{-0.016}^{+0.014}{}_{-0.023}^{+0.017}{}_{-0.003}^{+0.003}{}_{----}^{+---} $ \\[2mm] 
 \hline\hline 
\end{tabular}
\bigskip



\begin{tabular}{@{}crr@{}}
\hline\hline \\[-3mm] 
Observable $\av{F_L}$ & KMPW - scheme 1 \hspace{10mm} & BZ - scheme 1  \hspace{10mm} \\[2mm] 
 \hline \\[-3mm] 
$[0.1,0.98] $  & $0.213_{-0.016}^{+0.021}{}_{-0.137}^{+0.214}{}_{-0.033}^{+0.039}{}_{-0.006}^{+0.008}{}_{-0.043}^{+0.078} $ & $0.261_{-0.021}^{+0.024}{}_{-0.021}^{+0.025}{}_{-0.036}^{+0.040}{}_{-0.007}^{+0.009}{}_{----}^{+---} $ \\[2mm] 
 \hline \\[-3mm] 
$[1.1,2] $  & $0.646_{-0.025}^{+0.030}{}_{-0.267}^{+0.184}{}_{-0.049}^{+0.047}{}_{-0.008}^{+0.010}{}_{-0.039}^{+0.038} $ & $0.704_{-0.025}^{+0.027}{}_{-0.021}^{+0.023}{}_{-0.044}^{+0.038}{}_{-0.008}^{+0.009}{}_{----}^{+---} $ \\[2mm] 
 \hline \\[-3mm] 
$[2,3] $  & $0.761_{-0.015}^{+0.016}{}_{-0.244}^{+0.140}{}_{-0.029}^{+0.025}{}_{-0.004}^{+0.005}{}_{-0.025}^{+0.022} $ & $0.794_{-0.014}^{+0.015}{}_{-0.016}^{+0.017}{}_{-0.021}^{+0.020}{}_{-0.004}^{+0.004}{}_{----}^{+---} $ \\[2mm] 
 \hline \\[-3mm] 
$[3,4] $  & $0.770_{-0.007}^{+0.010}{}_{-0.249}^{+0.147}{}_{-0.016}^{+0.015}{}_{-0.003}^{+0.003}{}_{-0.017}^{+0.017} $ & $0.786_{-0.009}^{+0.010}{}_{-0.017}^{+0.018}{}_{-0.014}^{+0.014}{}_{-0.002}^{+0.002}{}_{----}^{+---} $ \\[2mm] 
 \hline \\[-3mm] 
$[4,5] $  & $0.739_{-0.004}^{+0.007}{}_{-0.265}^{+0.175}{}_{-0.013}^{+0.012}{}_{-0.002}^{+0.003}{}_{-0.021}^{+0.021} $ & $0.744_{-0.008}^{+0.009}{}_{-0.020}^{+0.021}{}_{-0.014}^{+0.013}{}_{-0.002}^{+0.002}{}_{----}^{+---} $ \\[2mm] 
 \hline \\[-3mm] 
$[5,6] $  & $0.697_{-0.005}^{+0.006}{}_{-0.275}^{+0.208}{}_{-0.014}^{+0.012}{}_{-0.003}^{+0.003}{}_{-0.029}^{+0.026} $ & $0.693_{-0.008}^{+0.009}{}_{-0.021}^{+0.023}{}_{-0.016}^{+0.015}{}_{-0.003}^{+0.003}{}_{----}^{+---} $ \\[2mm] 
 \hline \\[-3mm] 
$[6,7] $  & $0.654_{-0.005}^{+0.007}{}_{-0.276}^{+0.240}{}_{-0.014}^{+0.013}{}_{-0.004}^{+0.004}{}_{-0.037}^{+0.032} $ & $0.644_{-0.007}^{+0.009}{}_{-0.022}^{+0.025}{}_{-0.016}^{+0.015}{}_{-0.003}^{+0.004}{}_{----}^{+---} $ \\[2mm] 
 \hline \\[-3mm] 
$[7,8] $  & $0.613_{-0.005}^{+0.008}{}_{-0.271}^{+0.267}{}_{-0.013}^{+0.013}{}_{-0.005}^{+0.005}{}_{-0.051}^{+0.043} $ & $0.598_{-0.007}^{+0.010}{}_{-0.023}^{+0.025}{}_{-0.016}^{+0.015}{}_{-0.005}^{+0.005}{}_{----}^{+---} $ \\[2mm] 
 \hline \\[-3mm] 
$[1.1,2.5] $  & $0.681_{-0.023}^{+0.027}{}_{-0.263}^{+0.172}{}_{-0.044}^{+0.041}{}_{-0.007}^{+0.009}{}_{-0.035}^{+0.033} $ & $0.732_{-0.023}^{+0.024}{}_{-0.020}^{+0.021}{}_{-0.038}^{+0.033}{}_{-0.007}^{+0.008}{}_{----}^{+---} $ \\[2mm] 
 \hline \\[-3mm] 
$[2.5,4] $  & $0.771_{-0.009}^{+0.011}{}_{-0.247}^{+0.143}{}_{-0.019}^{+0.017}{}_{-0.003}^{+0.003}{}_{-0.019}^{+0.017} $ & $0.791_{-0.010}^{+0.010}{}_{-0.017}^{+0.018}{}_{-0.015}^{+0.015}{}_{-0.002}^{+0.002}{}_{----}^{+---} $ \\[2mm] 
 \hline \\[-3mm] 
$[4,6] $  & $0.717_{-0.005}^{+0.006}{}_{-0.271}^{+0.193}{}_{-0.014}^{+0.012}{}_{-0.003}^{+0.003}{}_{-0.025}^{+0.024} $ & $0.717_{-0.008}^{+0.009}{}_{-0.021}^{+0.022}{}_{-0.015}^{+0.014}{}_{-0.002}^{+0.003}{}_{----}^{+---} $ \\[2mm] 
 \hline \\[-3mm] 
$[6,8] $  & $0.632_{-0.005}^{+0.007}{}_{-0.274}^{+0.254}{}_{-0.013}^{+0.013}{}_{-0.004}^{+0.004}{}_{-0.045}^{+0.038} $ & $0.620_{-0.007}^{+0.010}{}_{-0.023}^{+0.025}{}_{-0.016}^{+0.015}{}_{-0.004}^{+0.004}{}_{----}^{+---} $ \\[2mm] 
 \hline\hline 
\end{tabular}
\bigskip
\end{document}